\documentclass[aps,preprintnumbers,amsmath,amssymb,prl,superscriptaddress,longbibliography,twocolumn]{revtex4-2}
\usepackage{mwe}
\usepackage{dcolumn}
\usepackage{color}
\usepackage{xcolor}
\usepackage[caption=false]{subfig}
\usepackage{feynmp}
\usepackage{feynmp-auto}
\usepackage{float}
\usepackage{systeme,mathtools}
\usepackage{easyReview}
\usepackage{amsmath,amssymb,amsfonts,empheq,physics,float,mathrsfs,wrapfig,indentfirst,enumitem}

\def\comment#1{}

\newcommand{\nc}{\newcommand}

\nc{\scs}{\scriptstyle}
\nc{\setval}{\fmfset{wiggly_len}{3mm} \fmfset{arrow_len}{1.5mm}
	\fmfset{arrow_ang}{13} \fmfset{dash_len}{1.5mm}\fmfpen{0.125mm}
	\fmfset{dot_size}{2thick}}

\usepackage{bm,latexsym,mathrsfs,enumerate,color}
\usepackage[mathcal]{euscript}
\usepackage[breaklinks=true,unicode=true,urlcolor = blue,colorlinks = true,citecolor = blue,linkcolor = blue]{hyperref}
\usepackage{graphicx}
\usepackage{todonotes}
\usepackage{wrapfig}
\usepackage{easyReview}
\usepackage{mathbbol}
\usepackage{stmaryrd}

\renewcommand{\vec}[1]{\bm{#1}}

\def\slashchar#1{\setbox0=\hbox{$#1$}           
	\dimen0=\wd0                                 
	\setbox1=\hbox{/} \dimen1=\wd1               
	\ifdim\dimen0>\dimen1                        
	\rlap{\hbox to \dimen0{\hfil/\hfil}}      
	#1                                        
	\else                                        
	\rlap{\hbox to \dimen1{\hfil$#1$\hfil}}   
	/                                         
	\fi}                                         %

\DeclareMathAlphabet\mathbfcal{OMS}{cmsy}{b}{n}
\DeclareSymbolFontAlphabet{\amsmathbb}{AMSb}%

\begin{document}
\title{Wilsonian Renormalization as a Quantum Channel and the Separability of Fixed Points}
\author{Matheus H. Martins Costa}
\affiliation{Institute for Theoretical Solid State Physics, IFW Dresden, Helmholtzstr. 20, 01069 Dresden, Germany}

\author{Jeroen van den Brink}
\affiliation{Institute for Theoretical Solid State Physics, IFW Dresden, Helmholtzstr. 20, 01069 Dresden, Germany}
\affiliation{Institute for Theoretical Physics and W\"urzburg-Dresden Cluster of Excellence ct.qmat, TU Dresden, 01069 Dresden, Germany}

\author{Flavio S. Nogueira}
\affiliation{Institute for Theoretical Solid State Physics, IFW Dresden, Helmholtzstr. 20, 01069 Dresden, Germany}

\author{Gast\~ ao I. Krein}
\affiliation{Instituto de F\'{i}sica Te\'orica, Universidade Estadual Paulista,
Rua Dr. Bento Teobaldo Ferraz, 271 - Bloco II, 01140-070 S\~ ao Paulo, Brazil}

\begin{abstract}
    We show that the Wilsonian formulation of the renormalization group (RG) defines a quantum channel acting on the momentum-space density matrices of a quantum field theory. This information theoretical property of the RG allows us to derive a remarkable consequence for the vacuum of theories at a fixed point: they have no entanglement between momentum scales. Our result can be understood as deriving from the scale symmetry of such theories and leads to constraints on the form of the ground state and on expectation values of momentum space operators. 
\end{abstract}

\maketitle

\textit{Introduction}.- The Wilsonian renormalization group (RG) transformation is a fundamental concept in the study of quantum field theories (QFTs) and statistical physics, which has been of great importance to understanding phenomena in these areas \cite{weinberg_1995,weinberg_1996,cardy_1996,goldenfeld}. It is typically defined as the integration over high momentum modes of a field theory above a given scale $\mu$ 
with a sharp cutoff, followed by the rescaling of momenta and fields \cite{wilson,Ma}

For physical quantum systems (whose states are described by density matrices \cite{nielsen}), it is of prime importance for the Wilson RG to be a quantum channel of states in a QFT, i.e., a completely positive and trace-preserving (CPTP) map (see Sec. 8.2 of Ref. \cite{nielsen} for a definition and discussion). Indeed, were it not CPTP, there would be field theories, possibly tensored with finite-dimensional systems, where the renormalization procedure gives rise to density matrices for the long distance degrees of freedom that would violate key requirements of quantum mechanics (e.g, positivity). Furthermore, since the RG preserves exactly the averages of long-wavelength observables \cite{Ma}, this would contradict the fact that the set of expectation values of all elements in an observable algebra determines the quantum state \cite{haag, pct}, and such expectation values are obtained from a vacuum state.

Although it is physically intuitive that the Wilsonian RG defines a CPTP map, such has so far not been demonstrated, despite recent advances discussed in Refs. \cite{wolf, casini2016, fliss, lashkari1,lashkari2, snowmass}. Our first result in this Letter is to prove that the RG has this property. This will be done via the Schrodinger picture of the wave functionals in QFT \cite{jackiw,fliss}. We then use it to explore entanglement properties between momentum scales at RG fixed points.

As discussed first in Refs. \cite{Balasubramanian:2011wt, Hsu:2012gk}, and later explicitly worked out with examples in Ref. \cite{paper1}, the first RG step, integrating out fast modes, is equivalent to taking a partial trace over degrees of freedom above a certain scale, 
which is a quantum channel. Therefore, we only need to focus on the rescaling of fields and momenta in the second step to show its CPTP property. 
We will show that the scaling used in the RG procedure defines a unitary evolution of the momentum-space density matrices. Hence, as the composition of two quantum channels is still CPTP, we conclude the proof for the full Wilsonian RG transformation. 
This nature of the RG not only guarantees the expected consistency of the method, but it also paves the way towards investigating how the entanglement between momentum scales evolves along the RG trajectory. In particular, it allows us to study such entanglement for theories lying at fixed points. Our main result is establishing that, as long as a fixed point theory exists (for example, as a CFT), then there is no entanglement in its ground state between momentum modes at different scales, even though such theories can be strongly interacting. We conclude with a discussion about some consequences of this property and on the novel perspectives it provides.

\textit{The Wilsonian RG as a quantum channel}.- We begin by reviewing one of the insights of Ref. \cite{Balasubramanian:2011wt}. It is known that the ground state of a quantum system can be represented via the functional integral of its action (see, for instance, Sec. IV in Ref. \cite{nishioka}). Suppose one partitions the Hilbert space of a QFT in $d$ spatial dimensions with action $S$ in momentum space as $\mathcal{H}=\bigotimes_{\vec{k}}\mathcal{H}_{\vec{k}}$, where each $\mathcal{H}_{\vec{k}}$ is generated by eigenstates of (the hermitian components of) the field mode $\phi_{\vec{k}}$, where $\phi$ represents any collection of bosonic and fermionic fields of the theory. Then, the ground state matrix elements between two vectors $\ket{\varphi_{\vec{k}}}$, $\ket{\Tilde{\varphi}_{\vec{k}}}$ such that each momentum mode has a definite amplitude are given by the path integral \cite{Balasubramanian:2011wt,nishioka},
\begin{equation}
    \bra{\varphi_{\Vec{k}}}\rho\ket{\Tilde{\varphi}_{\Vec{k}}} = \frac{1}{Z} \int_{\phi_{\Vec{k}}(0^-)=\Tilde{\varphi}_{\Vec{k}}}^{\phi_{\Vec{k}}(0^+)=\varphi_{\Vec{k}}}\mathcal{D}\phi_{\Vec{k}}(\tau) e^{-S[\phi_{\Vec{k}}(\tau)]},
\end{equation}
the boundary condition indicates a discontinuity at Euclidean time $\tau=0$, the action $S[\phi_{\Vec{k}}(\tau)]$ is written in terms of the Fourier-transformed fields and
\begin{equation}
    Z= \int\mathcal{D}\phi_{\Vec{k}}(\tau) e^{-S[\phi_{\Vec{k}}(\tau)]}.
\end{equation}
With this representation it becomes clear that integrating out fast modes with $|\vec{k}|\geq\mu$ for an arbitrarily chosen and changeable $\mu$, is the same as taking a partial trace over the Hilbert space $\otimes_{|\vec{k}|\geq\mu}\mathcal{H}_{\vec{k}}$, as can be seen by the equality

\begin{equation}
  \begin{split}
        \langle\mathbb{O}\rangle  &= \frac{1}{Z}\int\mathcal{D}\phi_{\Vec{k}}(\tau) \mathbb{O}\left(\phi_{\Vec{k}}, i\frac{\delta}{\delta\phi_{\Vec{k}}} \right)e^{-S[\phi_{\Vec{k}}(\tau)]}\\
        &= \frac{1}{Z}\int\mathcal{D}\phi_{|\Vec{k}|\leq\mu}(\tau) \mathbb{O}\left(\phi_{\Vec{k}}, i\frac{\delta}{\delta\phi_{\Vec{k}}} \right)e^{-S_\mu[\phi_{\Vec{k}}(\tau)]}
  \end{split}
\end{equation}
for any observable $\mathbb{O}$ built from field modes $\phi_{\Vec{k}}$ such that $|\Vec{k}|\leq \mu$ and where $S_\mu$ is the Wilsonian effective action at scale $\mu$ obtained by integrating out fields with $|\Vec{k}|>\mu$. This is the relation $\langle\mathbb{O}_A\rangle=\Tr(\rho\mathbb{O}_A)=\Tr_A(\rho_A\mathbb{O}_A)$ which characterizes a reduced density matrix for a subsystem $A$ from the observables acting on it, applied to momentum scales in a field theory, here defined as the field modes with momenta with a certain magnitude.

Thus, a low-momentum density matrix $\rho_\mu$ derived from this QFT ground state is well defined and given in terms of $S_\mu$ by \cite{Balasubramanian:2011wt,paper1},
\begin{equation}
    \bra{\varphi_{|\Vec{k}|<\mu}}\rho_\mu\ket{\Tilde{\varphi}_{|\Vec{k}|<\mu}} = \frac{1}{Z} \int_{\phi_{\Vec{k}}(0^-)=\Tilde{\varphi}_{|\Vec{k}|<\mu}}^{\phi_{\Vec{k}}(0^+)=\varphi_{|\Vec{k}|<\mu}}\mathcal{D}\phi_{\Vec{k}} e^{-S_\mu}
\end{equation}

The broader point is that this interpretation is valid even in the case of states other than the vacuum: the first RG step defines a partial trace over high-momentum modes and takes density matrices on the full Hilbert space of the theory to density matrices acting on the long-wavelength degrees of freedom only.

Moving on to the scaling transformation, we define $\Lambda$ as the overall cutoff of the QFT and the scaling parameter as $\sigma:=\Lambda/\mu$. Thus, the rescaling of field modes is given by \cite{wilson},
\begin{equation}
    k\to \sigma k,
\end{equation}
\begin{equation}
    \phi_{\Vec{k}} \to \sigma^{d_{\phi}} \phi_{\sigma\Vec{k}}, 
\end{equation}
where $d_\phi$ is the scaling dimension of the Fourier-transformed field, which depends on the fixed point of interest. The Euclidean time variable must also be rescaled as $\tau\to\sigma^{-z}\tau$, using the dynamical critical exponent $z$ introduced by Hertz in Ref. \cite{hertz}. We keep $z$ generic as our results will be valid for both relativistic and non-relativistic field theories. Furthermore, note that the scaling transformation employed here is simply the uniform dilation of length scales. More general Weyl transformations curving space are not investigated. The latter lead to anomalies in certain CFTs (the main differences between the two transformations are discussed in Ref. \cite{Nakayama}).

The matter of time rescaling is also a good opportunity to emphasize the peculiarities of our momentum-space cutoff: modes with $|\Vec{k}|>\mu$ are integrated over \textit{at all energies}, without any constraint in the temporal component of momentum, which transforms only under the second step of the RG. Such distinction is essential for the integration of fast modes to be identified with a partial trace, as the degrees of freedom of the system are labeled by the spatial momenta and in the path integral the dependence in $\tau$ is only used to project into the ground state, meaning there are no restrictions on its conjugate $k_0$. Similar conclusions, in the context of the functional RG, are reached in Refs. \cite{fliss, goldman2023exact} (which discuss the phase space and canonical structure) and Ref. \cite{cotler}. This suggests that even for relativistic theories, focusing only on the spatial momenta is key to understanding the entanglement properties of QFTs. 

Last but not least, as discussed by Hertz in Ref. \cite{hertz}, Sect. VI, and Millis in Ref. \cite{Millis}, the fixed points and universal quantities obtained with this cutoff are the same as in any other RG method, thus keeping the following analysis very general.

Now, recall that $S[\phi_{\Vec{k}}(\tau)]$ is the original action of the QFT in terms of momentum-space fields and let $S_{(\sigma)}[\phi_{\Vec{k}}(\tau)] := S_\mu[\sigma^{d_{\phi}} \phi_{\sigma\Vec{k}}(\sigma^{-z}\tau)]$ denote the new action at scale $\Lambda$ obtained from the scaling transformation. Then, by means of the path integral construction of matrix elements of a state operator, this action naturally defines the density matrix $\rho_{(\sigma)}$ via,
\begin{equation}
	\label{rescaling}
    \bra{\varphi_{\Vec{k}}}\rho_{(\sigma)}\ket{\Tilde{\varphi}_{\Vec{k}}} =\frac{1}{Z_{(\sigma)}} \int_{\phi_{\Vec{k}}(0^-)=\Tilde{\varphi}_{\Vec{k}}}^{\phi_{\Vec{k}}(0^+)=\varphi_{\Vec{k}}}\mathcal{D}\phi_{\Vec{k}} e^{-S_{(\sigma)}[\phi_{\Vec{k}}]},
\end{equation}
\begin{equation}
    Z_{(\sigma)} = \int\mathcal{D}\phi_{\Vec{k}}(\tau) e^{-S_{(\sigma)}[\phi_{\Vec{k}}]}.
\end{equation}

There is a priori no reason to believe that $\rho_{(\sigma)} = \ket{\Omega}\bra{\Omega}$, where $\ket{\Omega}$ is the ground state vector, since the action $S_{(\sigma)}[\phi_{\Vec{k}}]$ will be generally different from the original $S[\phi_{\Vec{k}}]$. The process of obtaining an effective action by integrating part of the momentum modes can be inverted by a scaling transformation only at an RG fixed point.

In general, the scaling transformation must be defined not only for $\rho_\mu$, but also for any density matrix acting on the low-momentum degrees of freedom. We will do so by using the Schrodinger representation of states in a QFT \cite{jackiw,fliss}, where a generic density matrix $\rho$ acting on the Hilbert space of momentum modes below scale $\mu$ can be formally written as the path integral
\begin{equation}
\label{general1}
	\rho = \int \prod_{|\vec{k}|,|\vec{k}'|\leq\mu}\mathcal{D}\phi_{\vec{k}}\mathcal{D}\phi'_{\vec{k}'}\rho(\phi_{\vec{k}},\phi'_{\vec{k}'})\ket{\phi_{\vec{k}}}\bra{\phi'_{\vec{k}'}},
\end{equation}
\begin{equation}
\label{normal1}
\int \prod_{|\vec{k}|\leq\mu}\mathcal{D}\phi_{\vec{k}}\rho(\phi_{\vec{k}},\phi_{\vec{k}})=1.
\end{equation}
Then, we define the scaling transformation as taking $\rho$ to a $\Tilde{\rho}$ such that,
\begin{equation}
\label{general2}
	\Tilde{\rho} =  \int \prod_{|\vec{k}|,|\vec{k}'|\leq\Lambda}\mathcal{D}\phi_{\vec{k}}\mathcal{D}\phi'_{\vec{k}'}\Tilde{\rho}(\phi_{\vec{k}},\phi'_{\vec{k}'})\ket{\phi_{\vec{k}}}\bra{\phi'_{\vec{k}'}},
\end{equation}
\begin{equation}
\label{element}
    \Tilde{\rho}(\phi_{\vec{k}},\phi'_{\vec{k}'})=\frac{1}{\mathcal{N}}\rho(\sigma^{d_{\phi}}\phi_{\sigma^{-1}\vec{k}},\sigma^{d_{\phi}}\phi'_{\sigma^{-1}\vec{k}'}),
\end{equation}
\begin{equation}
\label{normal2}
	\mathcal{N}=\int \prod_{|\vec{k}|\leq\Lambda}\mathcal{D}\phi_{\vec{k}}\rho(\sigma^{d_{\phi}}\phi_{\sigma^{-1}\vec{k}},\sigma^{d_{\phi}}\phi'_{\sigma^{-1}\vec{k}'}),
\end{equation}
which is composed of the same rescalings as before with a relabeling of the momentum modes. The normalizing factor $\mathcal{N}$ is introduced due to the scaling of fields in the path-integral measure.
This definition is exactly the same as Eq. (\ref{rescaling}) whenever the density matrix elements can be defined via an effective action, with $\mathcal{N}=Z_{(\sigma)}/Z$. This can be confirmed by writing Eq. (\ref{rescaling}) in the form of Eq. (\ref{element}) via a change of variables. Note that while the rescaling of momenta and fields enacts a shift in the labels of the degrees of freedom, the time rescaling by itself produces no change: in Eq. (\ref{rescaling}) the fields are integrated over all possible dependencies in $\tau$ (a consequence of no cutoffs being imposed on the energies) and the integration limits are taken at $\tau=0^{\pm}$, so the rescaling can be undone by a change of variables with no alterations in the final matrix elements. Interestingly, this is not the case at finite temperature, not studied in this paper, where the time periodicity is changed by the rescaling, see Ref. \cite{Millis} and the Supplemental Material.

From now on, it is important to define the theory in a box of volume $V$, an IR cutoff, so that the number of degrees of freedom is finite and the functional integrals and other quantities are well defined. With this cutoff the normalization constant becomes $\mathcal{N}\approx \sigma^{-d_\phi\mu^dV}$ as can be seen by comparing Eqs. (\ref{normal1}) and (\ref{normal2}) explicitly.

As discussed in Ref. \cite{Ma}, this scaling is ``trivial" in the sense that all statistical properties of the state at low-momentum degrees of freedom are preserved and all original expectation values can be recovered. In a quantum system this is tantamount to the transformation being described by a unitary map; in fact, if we define the operator,
\begin{equation}
    U \equiv \sqrt{\mathcal{N}}\int\prod_{|\vec{k}|\leq\mu}\mathcal{D}\phi_{\vec{k}}\ket{\sigma^{-d_\phi}\phi_{\sigma\vec{k}}}\bra{\phi_{\vec{k}}},
\end{equation}
by computing the necessary integrals with both UV and IR cutoffs, it is easy to show that given $U$ and density matrices of Eqs. (\ref{general1}) and (\ref{general2}), we have $\Tilde{\rho}=U\rho U^\dagger$. 
Furthermore, $U$ also obeys $UU^\dagger=U^\dagger U=\mathbb{1}$ and so the scaling transformation is indeed unitary. This can be tested, for example, by confirming the validity of the fact that the entropy of a density matrix is invariant under unitary transformations \cite{nielsen}. Indeed, it can be shown using the method and examples of Ref. \cite{paper1}, that the entropy of the matrices before and after scaling (at lowest nontrivial perturbative order) are equal \footnote{See the Supplemental Material for details on this point, which also includes Ref. \cite{Wegner_1974}}.

In real space the unitarity of scaling maps is well-known. What we have shown is that this property is also present in the specific transformation used in the momentum-space RG, which also includes scaling of field modes and time and which, although first defined only as a manipulation of the effective action, naturally leads to a map of density matrices. Therefore, the full Wilsonian RG procedure defines a quantum channel $\rho\to\rho_{(\sigma)}$ which is the composition of a partial trace over high-momenta (map $\rho\to\rho_\mu$) and a unitary induced by the rescaling operation (map $\rho_\mu\to\rho_{(\sigma)}$). The RG flow, being a Completely Positive and Trace-Preserving process, can thus be described using tools such as the operator-sum representation (see Chapter 8 of Ref. \cite{nielsen}).

\textit{Entanglement between scales at a fixed point}--- To see how the information theory formulation of the RG might be valuable, we apply it to study the momentum-space entanglement in the ground states of RG fixed points. By definition, these QFTs are such that $S^*_{(\sigma)}[\phi_{\Vec{k}}] = S^*[\phi_{\Vec{k}}]$ (the latter being the fixed point action, generally including all powers and derivatives of the field) no matter how many degrees of freedom are integrated over in the first step. Consequently, by the connection between action functionals and density operators explored earlier, we must have $\rho_{(\sigma)}=\rho = \ket{\Omega}\bra{\Omega}$, meaning $\rho_{(\sigma)}$ is a pure state and so $S_{EE}(\rho_{(\sigma)}) = 0$.
The entropy of interest is $S_{EE}(\rho_\mu) = -\Tr(\rho_\mu\log\rho_\mu)$, which gives the entanglement between low and high momenta. However, we showed that rescaling is a unitary (in this context only also the inverse of the partial trace), therefore
\begin{equation}
\label{result}
   S_{EE}(\rho_\mu) = S_{EE}(\rho_{(\sigma)}) = 0.
\end{equation}
Thus, the ground states of theories at an RG fixed point have no entanglement between different momentum scales. 

While all transformations were defined starting from a full regularization of the QFT, this does not restrict the validity of our result. We introduce the UV cutoff $\Lambda$ in order to regularize the theory, but scale-invariance makes its removal simple. Interactions are renormalized so that the theory is kept at the fixed point, by simply leaving the dimensionless parameters constant, and the limiting procedure $\Lambda\to\infty$ keeps the entanglement entropy between slow and fast modes equal to zero. As for the IR cutoff, Refs. \cite{cardy_1996,goldenfeld} explain how the finite volume acts as a relevant operator, driving the system away from the fixed point, and as pointed out in Ref. \cite{Ma}, the scaling transformation effectively changes the size of the box as $V\to\sigma^{-d}V$. This means that the $V\to\infty$ limit must be taken before any other when defining the theory, similarly to discussions of spontaneous symmetry breaking. This limit ensures the theory stays at a fixed point and that the Hilbert spaces before and after scaling are the same (without it, different periodic boundary conditions define different vector spaces), a necessary condition for the equation $\rho_{(\sigma)}=\ket{\Omega}\bra{\Omega}$ to be meaningful. Ultimately this does not change much, as the scaling transformation is still unitary at infinite volume and the proof of Eq. (\ref{result}) follows the same way, though it is important to keep these subtleties in mind.

Note that our only assumption at this point was that a fixed point exists. Hence, what we have shown is that, contrary to what may be intuitively expected, even a strongly interacting QFT can have no entanglement in its vacuum if the theory is scale-invariant. In other words, the stringent conditions on the couplings of a fixed point automatically constrain the ground state entanglement between momentum scales to vanish. 

Physically, this result can be understood as follows. The entanglement entropy between scales necessarily vanishes as both $\mu\to0$ and $\mu\to\Lambda$.  This is because as $\mu\to\Lambda$, then fewer and fewer momentum degrees of freedom are being integrated out, and in this limit we are simply left with the full vacuum of the theory, a pure state with zero entropy (when $\mu=\Lambda$ we simply have the full action of the theory, which defines the ground state). On the other hand, if $\mu\to0$, we can invoke the fact that if the global state is pure, the entropy after taking the partial trace is equal to that of the density matrix of the traced out degrees of freedom \cite{nielsen}, meaning that the entropy at $\mu=0$ must be equal to the entropy of the ground state, which is zero. 

Now, the entanglement entropy is always positive \cite{nishioka}, so it must reach a maximum between $0$ and $\Lambda$ as $\mu$ is varied, but such a maximum naturally defines a characteristic scale for a theory, since it is the momentum scale across which modes are correlated the most. Therefore, if a theory is scale-invariant, the momentum-space entanglement entropy must be constant. Since we know it vanishes both in the IR and UV extremes, it must vanish always. In this way we arrive once more at the conclusion that there must be no entanglement with respect to this partition. 
This general behavior of momentum-space entanglement used in our argument
is seen in the explicit formulas obtained in Refs. \cite{Balasubramanian:2011wt,paper1} and is the field theory equivalent of the ``Page curve" discussed in Sec. 3.1 of Ref. \cite{head}. The latter is an upper-bound on the entanglement entropy generated when degrees of freedom are gradually traced out in a pure state of a finite-dimensional system.

\textit{Consequences for fixed point theories} --- We can derive a number of implications from the fact that there is no entanglement between momentum scales in the ground state of scale-invariant QFTs for any separation scale $\mu$ chosen. 
The most direct one is that the vacua of these theories are separable, i.e., it is a simple tensor product of terms labeled by the momentum scale. Writing the Hilbert space of a fixed point theory as $\mathcal{H}=\bigotimes_\mu\mathcal{H}_\mu$, with $\mu$ denoting the momentum scale (meaning each $\mathcal{H}_\mu$ contains all modes with $|\vec{k}|=\mu$), the vacuum must be given by $\ket{\Omega}=\otimes_\mu\ket{\Omega_\mu}$. Due to scale symmetry and the unitarity of the scaling map, the projections of the components $\ket{\Omega_\mu}$ into eigenstates of field modes must obey $\bra{\phi_{\Vec{k}}}\ket{\Omega_\mu}=\bra{\sigma^{-d_\phi}\phi_{\sigma\Vec{k}}}\ket{\Omega_{\sigma\mu}}$ for any real $\sigma$.

Furthermore, separability of the state vector leads to connected correlation functions of observables acting on different momentum scales being all equal to zero \cite{nielsen,entropy}. That is, defining the operators which act on the subsystems below and above scale $\mu$, respectively

\begin{equation}
\label{observable1}
\mathbb{O}_{<} := \sum_{n=1}^\infty\int_{|\vec{k}_i|\leq\mu}\prod_{i=1}^n\frac{d^dk_i}{(2\pi)^d}f_n(\Vec{k}_1,...,\Vec{k}_n)\phi_{\Vec{k}_1}...\phi_{\Vec{k}_n}  
\end{equation}

\begin{equation}
\label{observable2}
\mathbb{O}_{>} := \sum_{n=1}^\infty\int_{|\vec{k}_i|>\mu}\prod_{i=1}^n\frac{d^dk_i}{(2\pi)^d}\Tilde{f}_n(\Vec{k}_1,...,\Vec{k}_n)\phi_{\Vec{k}_1}...\phi_{\Vec{k}_n}  
\end{equation}
given two families of functions $\{f_n(\Vec{k}_1,...,\Vec{k}_n)\}$, $\{\Tilde{f}_n(\Vec{k}_1,...,\Vec{k}_n)\}$ (which must be of compact support in $|\vec{k}_i|\leq\mu$, $|\vec{k}_i|>\mu$, respectively, see chapter 2 of Ref. \cite{haag}) then the separability of the vacuum implies the factorization of the expectation value of their product: $\langle\mathbb{O}_{<}\mathbb{O}_{>}\rangle=\langle\mathbb{O}_{<}\rangle\langle\mathbb{O}_{>}\rangle$.

Translating this condition into identities for the field correlators is somewhat complicated, but the $n$-point functions of the field in momentum space must be such that all momenta are at the same scale (have the same absolute value), or else they factorize into products of correlators. For example, the four-point function $\langle\phi_{\vec{k}_1}\phi_{\vec{k}_2}\phi_{\vec{k}_3}\phi_{\vec{k}_4}\rangle$ becomes such that
\begin{equation}
\begin{split}
\langle\phi_{\vec{k}_1}\phi_{\vec{k}_2}\phi_{\vec{k}_3}\phi_{\vec{k}_4}\rangle &= F(\vec{k}_1,\vec{k}_2,\vec{k}_3,\vec{k}_4) + \langle\phi_{\vec{k}_1}\phi_{\vec{k}_2}\rangle\langle\phi_{\vec{k}_3}\phi_{\vec{k}_4}\rangle \\
&+\langle\phi_{\vec{k}_1}\phi_{\vec{k}_3}\rangle\langle\phi_{\vec{k}_2}\phi_{\vec{k}_4}\rangle+\langle\phi_{\vec{k}_1}\phi_{\vec{k}_4}\rangle\langle\phi_{\vec{k}_2}\phi_{\vec{k}_3}\rangle,
\end{split}
\end{equation}
where $F(\vec{k}_1,\vec{k}_2,\vec{k}_3,\vec{k}_4)$ depends on the fixed-point theory and vanishes unless $|\vec{k}_1|=|\vec{k}_2|=|\vec{k}_3|=|\vec{k}_4|$. This identity can be understood as follows. If all momenta have same magnitude, the correlator can have any form consistent with scale symmetry, otherwise it must factorize into a product of expectation values.
Note that this result is independent of momentum conservation (the expression still contains a delta function making $\sum_{i=1}^4\vec{k}_i=0$). Furthermore, generalized versions of this relation are valid for the other $n$-point functions. 

It would be interesting to compare these formulas to the ones found in Refs. \cite{correlator1,correlator2} for CFTs, though the authors work with correlations of arbitrary scaling operators while we are considering the ``fundamental" field appearing in the Lagrangian defining the theory, in terms of which all operators may be constructed. Expanding on this latter notion, mathematically it means the field $\phi_{\Vec{k}}$ and its polynomials must define an irreducible set of operators in the Hilbert space of the QFT (see section 3.1 of Ref. \cite{pct} for an introduction). This requirement is what distinguishes operators whose momentum correlations must factorize between scales at a fixed point from the others: it formalizes the idea that a ``fundamental field" identifies the ``degrees of freedom" of a QFT. Going back to CFTs, a generic scaling operator does not satisfy this irreducibility condition and so its correlation functions do not have to factorize. 

Having discussed some corollaries of our result, it is important to make clear that the lack of entanglement between momentum scales does not imply that theories at an RG fixed point have an unentangled vacuum: the notion of entanglement depends on the chosen partition of the Hilbert space and separability with respect to one tensor product structure does not imply the same about other partitions. For example, in free field theories there is entanglement in real space but not in momentum space \cite{nishioka}.

\textit{Conclusions and Outlook} --- We have shown that the Wilsonian RG is equivalent to a quantum channel acting on density matrices of the momentum-space degrees of freedom. Furthermore, we proved that it is such that RG fixed points have no entanglement between momentum modes at different scales and discussed some of the consequences of this fact. 

The analysis made here can serve as starting point for other investigations, perhaps of QFTs at a phase transition instead of a fixed point. A field theory undergoing a second phase transition may still flow under the RG transformation, see Ref. \cite{graphene}. More broadly, we can use techniques such as the operator-sum decomposition to ask how specific RG flows reflect on the momentum space entanglement entropy: does it present ``critical scaling" under certain conditions? By plotting $S_{EE}(\rho_\mu)$ as a function of $\mu$, does the graph contain universal information? And what properties of a given phase transition or crossover can be read off from it? 

From a mathematical point of view, while we have used the Schr\"odinger picture following Refs. \cite{jackiw, fliss}, this was merely a way of representing the idea that low-momentum observables of a QFT can be constructed formally via functions of the Fourier-transformed fields, thus defining the momentum-space operators for each mode $\vec{k}$. By comparison with the local algebras of observables \cite{haag}, which have been important for studying entanglement in real-space \cite{entropy,pontello,holland,lin,nonab}, it would be interesting to rigorously and abstractly define the momentum-space algebras of observables and analyze their properties, possibly connecting with previous work in Refs.m\cite{Buchholz:1995gr,Lashkariscale}. In such formalism, the partial trace over fast modes becomes the restriction of the ground state to the subalgebra of low-momentum observables and the rescaling of fields and momenta translates into applying the dual map of the scaling unitary of density matrices to this subalgebra. Furthermore, while we considered fields at a fixed time in our arguments, it is known that in relativistic theories they are too singular \cite{haag}. An algebraic formulation would avoid this problem by considering observables acting at spatial momenta below a certain scale, but with arbitrary energy: the algebra associated with a ``cylinder" of radius $\mu$ in momentum space and infinitely extended along the energy axis. This not only corresponds to the partial trace over high-momentum degrees of freedom while avoiding ill-defined operators, but also makes clear that our subalgebra is invariant under the rescaling of time with a dynamical critical exponent, equivalent to what was previously discussed for density matrices. Another opportunity provided by this formulation is to investigate the connection between momentum-space entanglement and the effects of renormalization in real-space entanglement, such as those explored in Refs.  \cite{Iso0,Iso1,Iso2}.

We may also wonder what the separability in momentum space of the ground state of CFTs implies to holography. Finding the dual in AdS space of the momentum space density matrix $\rho_\mu$ is essential to tackling this question, but is an open problem as pointed out in Ref. \cite{Balasubramanian:2012hb}. Furthermore, it was shown in Ref. \cite{Grozdanov:2011aa} that the intuitive idea of restricting the AdS radial coordinate corresponds to a \textit{relativistic} Wilsonian cutoff, that is, the remaining modes must obey, in Euclidean signature, $k_0^2+\vec{k}^2\leq\Lambda$, a constraint on the energies which, as mentioned in Ref. \cite{Balasubramanian:2011wt} and discussed previously in this paper, is absent from the tensor product structure we are working with. Nevertheless, a proposal in Ref. \cite{pedraza} generalizes of the concept of entanglement wedge to momentum space and merits further investigation. Lastly, the description of the RG as a specific CPTP map possibly opens a path to connecting renormalization to recent discussions of circuit complexity in field theory, such as the ones in Refs. \cite{Jefferson:2017sdb,bhatta}, which have also been studied in relation to the AdS/CFT duality.

\begin{acknowledgments}
	
 We thank the Deutsche Forschungsgemeinschaft (DFG) for support through the W\"urzburg-Dresden Cluster of Excellence on Complexity and Topology in Quantum Matter – ct.qmat (EXC 2147, Project No. 39085490) and the Collaborative Research Center SFB 1143 (project-id 247310070). G.K was supported in part by Conselho Nacional de Desenvolvimento Cient\'{i}fico e Tecnol\' ogico - CNPq, Grant No. 309262/2019-4 and Funda\c{c}\~ ao de Amparo \`a Pesquisa do Estado de S\~ ao Paulo (FAPESP), Grant No. 2018/25225-9.
 
\end{acknowledgments}

\bibliography{citations.bib}

\appendix
\section{Supplemental Material}
\renewcommand{\theequation}{S.\arabic{equation}}

We prove the claim in the main paper, that the von Neumann entropies before and after the scaling transformation explicitly calculated at lowest non-trivial order using the method developed in \cite{paper1} are equal.

As a brief review, the method in Ref. \cite{paper1} consists of calculating the Rényi entropies $H_n(\rho)$ through the relation
\begin{equation}
    H_n(\rho) = \frac{1}{1-n}\lim_{\beta\to\infty}(\log Z_n(\beta)-n\log Z(\beta)),
\end{equation}
where, given an effective action $S_{eff}$ which generates the matrix elements of $\rho$, $Z(\beta)$ is the usual finite-temperature partition function and $Z_n(\beta)$ is the partition function after modifying the non-local kernels of $S_{eff}$ in a specific manner detailed in Ref. \cite{paper1}.

It turns out that when starting with a free field theory and adding a perturbative interaction, a series of cancellations happen and at order $\mathcal{O}(\lambda^2)$ in the coupling (the lowest with non-trivial results) the von Neumann and Rényi entropies are proportional to the same contractions of Feynman diagrams appearing in the modified partition function.

Then, to show that the entropies before and after scaling are the same, we need only to prove the equality between Feynman diagram contractions. We will do so for one of the contributions, as the others follow the same argument.

Consider a contributing term to the entropy of reduced density matrix $\rho_\mu$ in perturbative $\lambda\phi^4$ theory of the form
\begin{equation}
\int^* \mathcal{K}_{\mu,\beta}(\Vec{k},\Vec{p},\Vec{q};\tau,\tau')\langle\phi_{\Vec{k}}(\tau)\phi^*_{\Vec{k}}(\tau')\rangle_{n\beta}
\end{equation}
corresponding to Eq. (C6) of Ref. \cite{paper1}, where the subscript $n\beta$ in the correlator means that the expectation value is taken at inverse temperature $n\beta$, the region of integration over all momenta and form of the kernel $\mathcal{K_{\mu,\beta}}$ are specified but irrelevant to our argument and a number of Matsubara sums and Euclidean time integrals are suppressed.

Applying the scaling map $\Vec{k}\to\sigma\Vec{k}$, $\phi_{\Vec{k}}\to \sigma^{d_\phi}\phi_{\sigma\Vec{k}}$, $\tau\to\sigma^{-z}\tau$, the associated term leading to the entropy of state $\rho_{(\sigma)}$ is
\begin{equation}
\label{new}
\begin{split}
    \int^*\mathcal{K}_{\mu,\beta}(\sigma^{-1}\Vec{k}',\Vec{p},\Vec{q};\sigma^{-z}\tau,\sigma^{-z}\tau')\\
\times\sigma^{1-d}\sigma^{2d_\phi}\langle\phi_{\Vec{k}'}(\sigma^{-z}\tau)\phi^*_{\Vec{k}'}(\sigma^{-z}\tau')\rangle_{n\beta}.
\end{split}
\end{equation}
Now, this transformation is defined such that $\langle\phi_{\Vec{k}}(\tau)\phi^*_{\Vec{k}}(\tau')\rangle_{n\beta} = \sigma^{2d_\phi}\langle\phi_{\sigma\Vec{k}}(\tau)\phi^*_{\sigma\Vec{k}}(\tau')\rangle_{n\beta}$ for the transformed fields, see Ref. \cite{Ma}. To deal with the rescaling in time we make a change of variables to restore $\tau, \tau'$, but as pointed out by Ref. \cite{Millis} and can be seen by taking into account the integration limits of the (suppressed) time integrals, this effectively changes the temperature periodicity to $\sigma^{-z}\beta$ (and $n\sigma^{-z}\beta$ in the replica trick calculations of Ref. \cite{paper1}). Therefore, it is easy to see that Eq. \eqref{new} equals to
\begin{equation}
\begin{split}
  \int^*\mathcal{K}_{\mu,\sigma^{-z}\beta}(\sigma^{-1}\Vec{k}',\Vec{p},\Vec{q};\tau,\tau') \times\\
    \sigma^{1-d}\langle\phi_{\sigma^{-1}\Vec{k}'}(\tau)\phi^*_{\sigma^{-1}\Vec{k}'}(\tau')\rangle_{n\sigma^{-z}\beta} =
    \\ \int^* \mathcal{K}_{\mu,\sigma^{-z}\beta}(\Vec{k},\Vec{p},\Vec{q};\tau,\tau')\langle\phi_{\Vec{k}}(\tau)\phi^*_{\Vec{k}}(\tau')\rangle_{n\sigma^{-z}\beta}
\end{split}
\end{equation}
where the last equality is derived via a simple change of variables in the momentum $\Vec{k}'$, originally one of the slow modes.

So we can see that there is a change for any \textbf{finite temperature} calculation, which makes an analogous investigation of the RG in this context an interesting problem. For our focus on the vacua of field theories at zero temperature, however, this is not a concern because the $\beta\to\infty$ limit remains unchanged and after the limit the results are the same as before the dilation. Thus, the contribution to the entropy of $\rho_{(\sigma)}$ is exactly equal to that of $\rho_{\mu}$ and a calculation can be done for any of the other perturbative terms leading to similar results. Therefore the total entropy is unchanged, consistent with our claim of the unitarity of the scaling transformation.

Finally, note that there was no need to specify the values of $d_\phi$ or $z$; the scaling is unitary regardless of the dimension given to field $\phi_{\Vec{k}}$. For the appearance of scaling dimensions different from the correct ones in the context of the renormalization group, see Ref. \cite{Wegner_1974}. In more detail, any scaling with wrong dimension can be decomposed into a product of the correct scaling with a change of the normalization of the field operator, which Ref. \cite{Wegner_1974} names a ``redundant operation".
\end{document}